\PassOptionsToPackage{table,xcdraw}{xcolor}
\documentclass[sigconf,nonacm]{acmart}

\usepackage{booktabs}
\usepackage{threeparttable}
\usepackage{tabularx}
\usepackage{caption}
\AtBeginDocument{%
  }

\setcopyright{acmlicensed}
\copyrightyear{2027}
\acmYear{2027}
\acmDOI{XXXXXXX.XXXXXXX}
\acmConference[ASP-DAC '27]{32th Asia and South Pacific Design Automation Conference}{Jan. 25-28, 2027}{Tokyo, Japan}
\acmISBN{XXX-X-XXXX-XXXX-X/2027/01}

\begin{document}

\title{LLMET: Enabling Cross-Layer Evaluation of Emerging M3D Memories for Energy-Efficient LLM Serving}


\author{Ming-Yen Lee$^{*1}$, Hanchen Yang$^{*1}$, Faaiq Waqar$^{1}$, Harsono Simka$^{2}$, Tushar Krishna$^{1}$, Muhammed Ahosan Ul Karim$^{2}$, Shimeng Yu$^{1}$}
\affiliation{%
  \institution{}
  \country{$^{1}$Georgia Institute of Technology, Atlanta, GA, USA \quad $^{2}$Samsung Semiconductor Inc., San Jose, CA, USA}
}
\email{mlee838@gatech.edu, hanchen@gatech.edu, shimeng.yu@ece.gatech.edu}


\renewcommand{\shortauthors}{M.-Y. Lee et al.}

\begin{abstract}
  The energy consumption of Large Language Model (LLM) serving is becoming a major system challenge as deployment scales, driven by hardware power and thermal constraints and rising electricity costs. A key contributor to chip energy dissipation is data movement between limited on-chip cache and off-chip High Bandwidth Memory (HBM). Meanwhile, emerging memory technologies such as monolithic 3D (M3D) integration of cache memories at the Back-End-Of-Line (BEOL) of logic chips enable larger and denser on-chip memories, creating new opportunities to reduce costly off-chip traffic. However, it remains unclear whether continuously scaling on-chip memory using emerging technologies can effectively improve the energy efficiency of LLM serving. To address this gap, we develop LLMET (\underline{LLM} with \underline{E}merging \underline{T}echnology), a validated cross-layer simulation framework, and conduct a comprehensive study on the impact of large-capacity on-chip memory technologies across a broad range of models, applications and platforms. Utilizing M3D technology to expand the L2 cache from 40MB to 1GB yields a 44 \% reduction in chip energy during the Llama3.1-70B prefill phase with a 16K context window, based on LLMET simulation on a dual NVIDIA A100 GPU setup. On the 8xNVIDIA B200-like platform, extending the L2 cache from 128MB to 4GB saves the prefill energy by up to 24\%. For the edge platform and workloads, the decode energy saving reaches 30 \% when increasing the 8MB cache size to 256MB. These results highlight the promise of ultra-large on-chip memories for energy-efficient LLM serving systems.
\end{abstract}



\maketitle
\begingroup\renewcommand\thefootnote{}\footnotetext{$^{*}$Both authors contributed equally to this research.}\addtocounter{footnote}{-1}\endgroup
\vspace{-6pt}

\section{Introduction}
The rapid growth of large language models (LLMs)~\cite{attention2017, gpt3, llama3} is placing increasing pressure on the memory subsystem, where frequent transfers of heavy model weights and long KV caches between on-chip cache and High Bandwidth Memory (HBM) dominates the energy consumption of AI accelerators~\cite{aimemorywall, hbmenergy, pawlowski_memory2019, horowitz}. Meanwhile, emerging memory technologies are expanding the design space of on-chip storage for energy-efficient LLM systems. Monolithic 3D (M3D) integration enables high-density, low-latency, and energy-efficient embedded memories at the Back-End-Of-Line (BEOL) of logic chips. Recent foundry proposals based on amorphous oxide semiconductor (AOS) devices demonstrate eDRAM-like buffers with capacities potentially reaching hundreds of megabytes \cite{samsungigzo, tsmc1t1c}. Previous studies have extensively explored these technologies from the perspective of device, circuit, and micro-architecture, demonstrating their potential for general-purpose computing platforms \cite{nscache, cmos+x}. These advances therefore raise a critical question of whether on-chip memory scaling with advanced memory technology can meaningfully address the memory bottleneck in LLM serving. In particular, \emph{Can Emerging M3D Memories Enable Energy-Efficient LLM Serving?}

\begin{table*}[t]
\centering
\caption{Comparison of LLMET with representative LLM modeling tools and simulators.
}
\label{tab:simulator_comparison}
\vspace{-4pt}
\resizebox{\textwidth}{!}{
\begin{tabular}{lccccc}
\toprule
\textbf{Work / Tool} &
\textbf{Emerging Memory /} &
\textbf{Detailed Memory-} &
\textbf{Cache-Aware} &
\textbf{Energy/PPA} &
\textbf{Main Scope} \\
&
\textbf{M3D Support} &
\textbf{Hierarchy Traffic} &
\textbf{Mapping/Fusion} &
\textbf{Breakdown} &
\\
\midrule
Calculon~\cite{calculon}
& No
& No
& No
& No
& High-level LLM system co-design \\

Vidur~\cite{vidur}
& No
& No
& No
& No
& LLM serving performance simulation \\

GenZ~\cite{genz}
& No
& No; capacity/BW only
& No
& No
& Platform requirement analysis \\

LLMCompass~\cite{llmcompass}
& No
& Yes
& Mapping only
& Area/cost only
& LLM hardware design exploration \\

DynamoLLM~\cite{dynamollm}
& No
& No
& No
& No; profile-based energy management
& Cluster-level energy optimization \\

LLMServingSim 2.0~\cite{llmservingsim2}
& Profile-based only
& System-level/profile-based
& No
& Power/profile metrics only
& Heterogeneous/disaggregated serving simulation \\

\midrule
\textbf{LLMET (This Work)}
& \textbf{Yes}
& \textbf{Yes}
& \textbf{Yes}
& \textbf{Yes}
& \textbf{Cross-layer simulation for LLM serving system w/ emerging technology support.} \\
\bottomrule
\end{tabular}
}
\end{table*}

Prior work has studied LLM serving from several perspectives~\cite{vllm, orca, distserve, splitwise, sarathi, streamingllm, h2o}. Large on-chip buffers have been used for KV-cache storage and prefetching to reduce latency and HBM bandwidth pressure \cite{llmtimeloop}, but their energy impact remains unclear. Cluster-level studies optimize throughput and performance across multi-GPU systems \cite{dynamollm}, yet do not capture per-chip memory-access energy. Several simulators provide fast performance modeling for LLM inference \cite{llmtimeloop,llmcompass,genz,vidur}, and some explore memory hierarchy scaling \cite{llmcompass}. Despite the importance of understanding the cross-layer impact of emerging memory technologies on LLM serving platforms, existing tools and prior studies do not capture these insights.

\begin{figure}[t]
\centerline{\includegraphics[width=19.5pc]{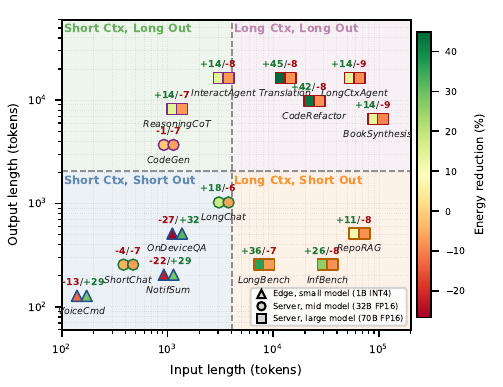}}
\caption{Impact of large on-chip memory across LLM applications. Each workload is plotted as a pair of markers (left = prefill, right = decode) and annotated as ``$\pm$P/$\pm$D'', where the first value is the prefill and the second is the decode total-energy reduction (\%) obtained by scaling the L2 cache from 8MB to 256MB for edge and from 32MB to 1024MB for server. Positive values denote energy savings; negative values denote overhead. Workloads span representative \emph{edge} use cases (Voice/Command, Notification Summary, On-device QA on Llama-3.2 1B INT4) and \emph{server} inference (Llama-3.1 32B and 70B in FP16) covering interactive chat, long-context QA (e.g., LongBench, InfBench), code generation and refactoring, multi-turn agentic dialog, and long-doc synthesis, with each workload modeled by its (input, output) token length.}\vspace*{-15pt}
\label{intro}
\end{figure}

To address these limitations, we make the following contributions:
\vspace{-6pt}
\begin{itemize}
    \item We present LLMET,
    a cross-layer simulation framework for LLM serving that supports emerging memory technologies. LLMET combines front-end execution tracing with back-end power, performance, and area (PPA) modeling to evaluate custom accelerators with large on-chip memories enabled by emerging M3D memory technology.

    \item Using LLMET, we conduct a comprehensive study on the energy-efficiency of LLM serving across models, applications, platforms, and emerging technologies. Figure~\ref{intro} shows a snapshot of the scope of our study.
    
    \item We validate LLMET against publicly reported NVIDIA A100 data \cite{llmcompass,a100datasheet}. The estimation error for both total die area and core area is within 7\%.

    \item Guided by the results, we provide design insights and highlight future directions and use cases for emerging memory technologies.
\end{itemize}

\section{Background and Related Work}
\subsection{LLM Serving Architecture}
Modern LLMs are built on the decoder-only transformer architecture~\cite{attention2017, llama3}, in which each layer interleaves multi-head attention with a feed-forward network (FFN). Serving an inference request proceeds in two distinct phases~\cite{scalingtransformer, vllm}. In the \emph{prefill} phase, the entire input prompt is processed in parallel to populate the key-value (KV) cache and emit the first output token; this phase is dominated by large general matrix multiplications (GEMMs) and is typically compute-bound. In the \emph{decode} phase, output tokens are generated autoregressively one at a time, with each step reloading the full model weights and the growing KV cache from memory to compute a single token; the matrix-vector nature of this phase makes it strongly memory-bound~\cite{genz,llmcompass}. The KV cache itself scales linearly with batch size and sequence length, so long-context and high-throughput serving place increasing pressure on the memory subsystem~\cite{vllm}.

Because on-chip cache capacity (tens of MB on current accelerators) is far smaller than the multi-gigabyte footprint of model weights and KV cache, these tensors are repeatedly streamed between High Bandwidth Memory (HBM) and the compute cores. This off-chip data movement, rather than arithmetic, dominates both latency and energy: an HBM access costs roughly two orders of magnitude more energy than an on-chip SRAM access~\cite{aimemorywall,hbmenergy,pawlowski_memory2019}. Enlarging on-chip memory so that more weights and KV data can be retained and reused on chip directly reduces this costly traffic~\cite{llmtimeloop}, which is the central opportunity that motivates this work.

\subsection{Emerging M3D Technology}
\begin{figure}[t]
\centerline{\includegraphics[width=20.0pc]{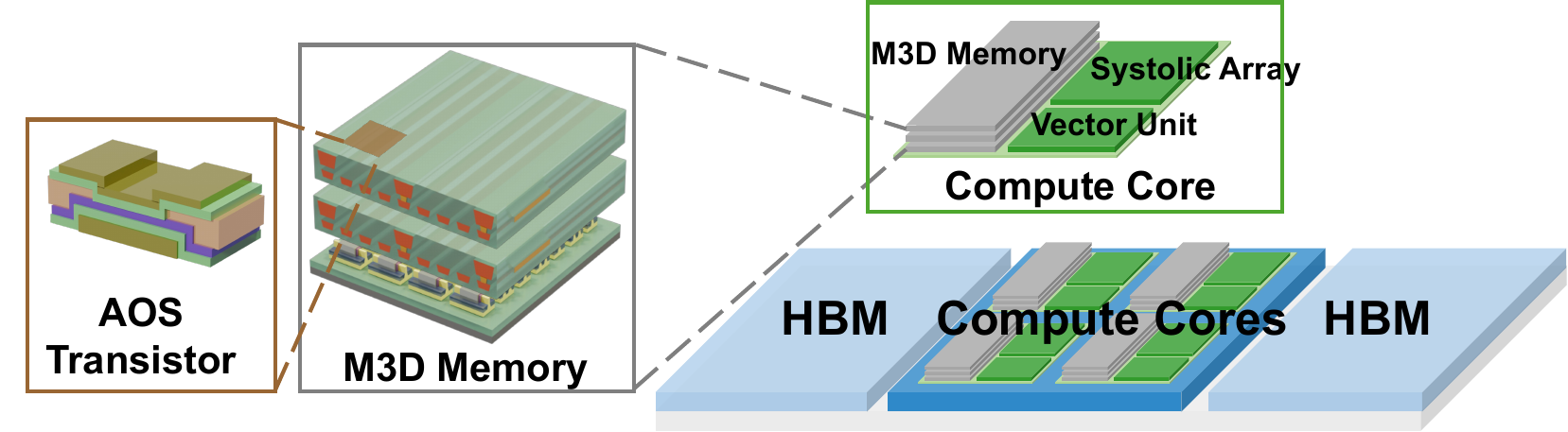}}
\caption{AI accelerators with large Monolithic 3D (M3D) on-chip memories.}\vspace*{-15pt}
\label{background}
\end{figure}

Monolithic 3D (M3D) embedded memories offer a scalable and high-density alternative for implementing next-generation, ultra-large on-chip caches. Driven by recent manufacturing milestones from leading foundries \cite{samsungigzo, tsmc1t1c}, amorphous oxide semiconductor (AOS) transistors enable low-leakage, high-speed embedded DRAM (eDRAM) buffers. This "CMOS+X" architecture shown in Figure \ref{background} integrates dense memory cells, such as two-transistor gain cell (2T-GC), directly within the back-end-of-line (BEOL) interconnects, vertically stacked above the front-end-of-line (FEOL) logic and peripheral circuits. By bypassing traditional 2D planar constraints, this approach can scale on-chip memory capacity to several hundred megabytes \cite{cmos+x}, offering the space for data storage and reuse.

State-of-the-art commercial implementations like AMD's 3D V-Cache achieve a 96MB cache capacity by bonding a separate SRAM die on top of the microprocessor core \cite{loh3d, amdvcache}. However, vertically stacking multiple SRAM layers to achieve larger capacities remains severely constrained by thermal dissipation issues, aggravated by the dozens-of-micrometers thickness of the bonded dies. On the other hand, M3D architectures are fabricated directly within the standard BEOL metallization stack (spanning only a few micrometers). This ultra-thin profile yields superior thermal characteristics, tighter interconnect pitches, and lower data-transfer latency.

\subsection{LLM Modeling Tools and Simulators}
A growing body of tools models LLM inference at different levels of abstraction, but none capture the cross-layer interaction between emerging on-chip memory technologies and serving energy that this work targets. At the cluster and serving level, Vidur~\cite{vidur} and LLMServingSim~2.0~\cite{llmservingsim2} simulate request scheduling and throughput across heterogeneous or disaggregated platforms, while DynamoLLM~\cite{dynamollm} optimizes cluster-level energy through profile-based power management; none of these expose per-chip memory-access energy or the on-chip memory hierarchy. At the system co-design level, Calculon~\cite{calculon} and GenZ~\cite{genz} provide fast high-level performance and platform-requirement analysis, but treat memory only as aggregate capacity and bandwidth rather than modeling hierarchical traffic. Closest to our work, LLMCompass~\cite{llmcompass} performs detailed hardware design exploration with memory-hierarchy traffic and area modeling, yet it assumes a fixed tiling strategy and reports only area and cost, leaving it unable to expose how off-chip traffic scales as on-chip capacity grows.

Critically, all of these tools assume conventional SRAM/HBM memory and lack support for emerging M3D memory technologies, cache-aware mapping and operator fusion, and a per-component energy breakdown---the three capabilities required to evaluate ultra-large on-chip memories for LLM serving. Table~\ref{tab:simulator_comparison} summarizes this gap, and LLMET is designed to fill it; we detail the framework and its distinguishing features in Section~\ref{sec:framework}.

\section{Proposed LLMET Framework}
\label{sec:framework}
\begin{figure*}[t!]
\centerline{\includegraphics[width=42.0pc]{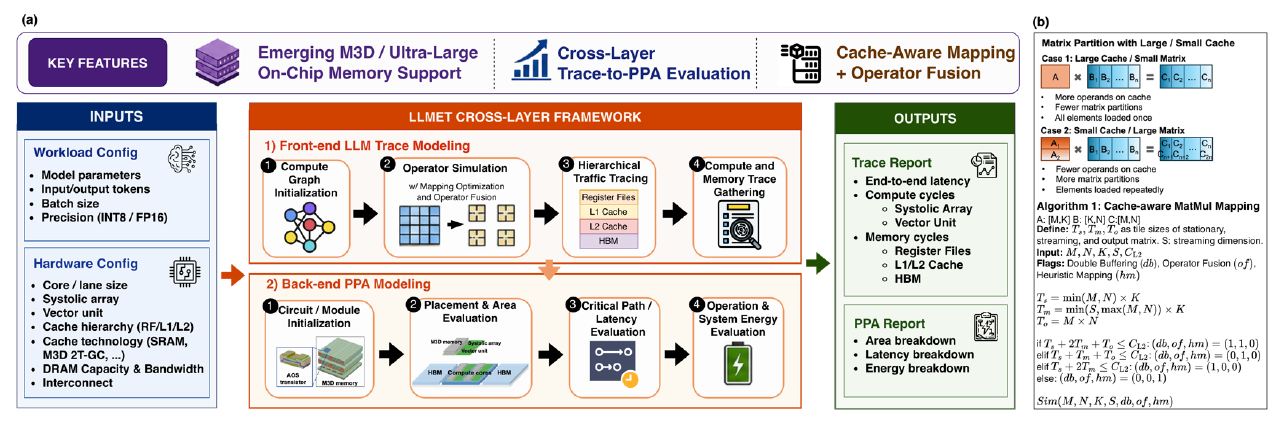}}
\captionsetup{skip=5pt}
\caption{(a) LLMET overview. (b) Cache-aware mapping in LLMET.}\vspace*{-10pt}
\label{llmet}
\end{figure*}

In this section, we present LLMET, a cross-layer evaluation framework for LLM inference workloads, with three key features that distinguish LLMET from prior LLM simulators (Table~\ref{tab:simulator_comparison}). 
First, \emph{device-calibrated, technology-specific PPA}: per-bit access energy and area for emerging M3D 2T gain-cell (2T-GC) memories are drawn from circuit-level NS-Cache models~\cite{nscache}, rather than the profile-based abstractions used in LLMServingSim~2.0~\cite{llmservingsim2}, enabling forward-looking evaluation of memory technologies that have not yet been fabricated at full capacity. 
Second, \emph{capacity-aware mapping}: LLMET selects from four operator-fusion regimes per operator as a function of available L2, exposing how off-chip traffic scales with cache size---an effect that fixed-tiling tools such as LLMCompass~\cite{llmcompass} and GenZ~\cite{genz} cannot capture. 
Third, \emph{hardware-component-level energy accounting}: coupled with the device-level backend, LLMET emits a per-component energy breakdown (DRAM, L2/L1/RF, systolic array, vector unit, on-chip interconnect) suitable for cross-layer co-design, rather than the aggregate latency or cluster-level power numbers reported by Calculon~\cite{calculon}, Vidur~\cite{vidur}, and DynamoLLM~\cite{dynamollm}. The backend hardware calibration is validated against published A100 die-area within 7\% (Figure~\ref{eval4}). Such implementation and integration enable LLMET to precisely model the cross-layer effects of memory technologies on LLM applications.

\subsection{Cross-Layer Modeling}
As shown in Figure~\ref{llmet}(a), LLMET takes both LLM model parameters and a hardware configuration as inputs. The front-end produces a per-operator trace that carries (i) the operator type and tensor shapes; (ii) the tile sizes selected by the mapping pass; (iii) the chosen capacity-aware mapping case (Section~B); (iv) the byte counts read from and written to each level of the memory hierarchy (RF, L1, L2, DRAM, link); and (v) compute-cycle counts per functional unit; the back-end consumes that trace, together with an instantiated device model, to evaluate system PPA. 

\subsection{Capacity-Aware Mapping + Operator Fusion}
The front-end uses cache-aware mapping and inter-operator fusion~\cite{timeloop} to expose the benefit of ultra-large on-chip caches (Figure~\ref{llmet}(b)). For matrix multiplication, if the L2 can hold one input operand, the smaller matrix is pinned on chip as a static operand while the other is streamed in tiles from DRAM. The remaining capacity then determines which of four mapping regimes is selected (Algorithm 1 in Figure~\ref{llmet}(b)). This capacity-aware case selection is what lets LLMET expose how off-chip traffic scales with on-chip cache size---an effect that fixed-tiling simulators, which assume a single mapping regardless of cache, cannot capture.

To precisely capture modern LLM system behavior, LLMET also performs inter-operator fusion in attention layers~\cite{flashattention, flashattention2}. The $\mathrm{QK^{T}}$ and $\mathrm{SV}$ stages are scheduled head-by-head, so the only DRAM traffic per head is loading $Q, K, V$ and writing the attention output $A$, while intermediate logits and softmax results stay on chip. This fusion is profitable when the per-head working set $n_{\text{seq}} \cdot d_{\text{head}} \cdot \text{bytes}\cdot(|Q|+|K|+|V|+|A|)$ fits in the cache. For grouped-query attention (GQA)~\cite{gqa, mqa}, the reused KV heads are pinned across all query heads in a group whenever $C_{L2} \ge n_{\text{kv}} \cdot n_{\text{seq}} \cdot d_{\text{head}} \cdot \text{bytes}$, eliminating redundant KV reloads across the entire group---a regime that becomes critical for $\ge$16K-token contexts where attention KV dominates DRAM traffic.

\subsection{Device-Calibrated PPA with Emerging-Memory Support}
The back-end instantiates hardware modules from the target architecture and evaluates each with a technology-calibrated model. Compute units (systolic arrays, vector units, DFF chains)~\cite{tpu, eyeriss} are derived from ASAP7 RTL synthesis~\cite{asap7}. SRAM components (RF, L1, L2) are evaluated with NS-Cache~\cite{nscache}. M3D 2T-GC memories are evaluated with the same NS-Cache flow but using the 2T-GC device model and the 3D layer-stacking area cost rather than 2D H-tree area, so technology changes propagate directly into per-bit energy and area. On-chip interconnects are modeled with NeuroSim~\cite{neurosim}, and HBM/LPDDR I/O energy is taken from foundry-reported per-bit numbers~\cite{hbmenergy}. 
This front-end/back-end co-design framework 
makes LLMET \textbf{the first} LLM system simulator that can propagate device-level design choices all the way to system-level serving energy and performance, 
turning emerging-memory pathfinding for LLM and AI accelerators from a device-only debate into a measurable system-design choice. Next we present our exploration and findings with LLMET.

\section{Evaluations}
\label{sec:eval}

Using LLMET, we study three deployment regimes that span server, technology-scaling, and edge LLM serving. The \emph{server} study uses Llama 3.1 70B on a 2$\times$NVIDIA A100 platform at 7nm technology node~\cite{a100datasheet}. The \emph{technology-scaling} study uses Llama 3.1 405B on an 8$\times$NVIDIA B200-like platform at 3nm technology node, with parameters extrapolated from disclosed Blackwell performance metrics~\cite{b200datasheet1,b200datasheet2} relative to the A100 baseline. The \emph{edge} study uses Llama 3.2 1B (INT4) on an edge platform following the hardware configurations of Jetson Orin NX-class accelerators \cite{jetsonorin} with LPDDR5 main memory at a 7nm technology node. The 7nm platform is benchmarked via ASAP7 RTL synthesis \cite{asap7}, whereas the 3nm node utilizes projection values from NeuroSim \cite{neurosim}, which are aligned with the 2024 International Roadmap for Devices and Systems (IRDS) \cite{irds2024}. Within each regime, we sweep the on-chip L2 cache capacity from a small SRAM baseline up to GB-scale M3D memory and report HBM traffic, total inference energy (compute and data movement), and chip-area overhead. Sweep ranges per regime are specified in the corresponding subsections.

\begin{table}[h]
\centering
\renewcommand{\tabularxcolumn}[1]{m{#1}}
\begin{threeparttable}
\caption{Memory Access Energy}
\label{table_energy}
\vspace{-8pt}
\renewcommand{\arraystretch}{0.5}
\begin{tabularx}{\linewidth}{
>{\scriptsize\hsize=0.45\hsize\arraybackslash}X
>{\scriptsize\hsize=0.4\hsize\arraybackslash}X
>{\scriptsize\hsize=0.4\hsize\arraybackslash}X
>{\scriptsize\hsize=0.45\hsize\arraybackslash}X
>{\scriptsize\hsize=0.4\hsize\arraybackslash}X
>{\scriptsize\hsize=0.4\hsize\arraybackslash}X
>{\scriptsize\hsize=0.45\hsize\arraybackslash}X
>{\scriptsize\hsize=0.45\hsize\arraybackslash}X}
\toprule
Memory & 7nm 40MB SRAM & 7nm 1GB SRAM & 7nm 1GB 2T-GC & 3nm 128MB SRAM & 3nm 1GB SRAM & 3nm 1GB 2T-GC & 3nm 4GB 2T-GC\\ 
\midrule
Access Energy \newline (pJ/bit) & 0.495 & 2.78 & 1.08 & 0.917 & 2.42 & 0.768 & 0.981 \\ 
\midrule
Memory & \multicolumn{2}{c}{\scriptsize HBM2E} & \multicolumn{2}{c}{\scriptsize HBM3E} & \multicolumn{3}{c}{\scriptsize LPDDR5} \\
 \midrule
Access Energy \newline (pJ/bit) & \multicolumn{2}{c}{\scriptsize 6.6} & \multicolumn{2}{c}{\scriptsize 5.74} & \multicolumn{3}{c}{\scriptsize 6.6} \\
\bottomrule
\end{tabularx}

\begin{tablenotes}
\scriptsize
    \item * HBM access energy obtained from \cite{hbmenergy}. IO energy included.
    \item * HBM2E and 7nm cache are implemented in A100 platform, while HBM3E and 3nm cache are implemented in B200-like platform.
    \item * LPDDR5 access energy is modeled to be the same as HBM2E \cite{pawlowski_memory2019}.
\end{tablenotes}
\end{threeparttable}
\end{table}

It should be noted that expanding cache with planar SRAM incurs prohibitive H-tree routing overhead at GB-scale capacities, while M3D 2T-GC delivers comparable per-bit access energy at a much smaller area footprint~\cite{cmos+x}. Therefore, we adopt M3D 2T gain-cell (2T-GC) memory at 128MB per layer as the default ultra-large cache. The 2D SRAM baseline is sized to each reference platform---40MB for the A100, 128MB for the B200-like configuration, and 8MB for the edge platform. Per-bit access energies for DRAM, planar SRAM, and 2T-GC at the relevant technology nodes are summarized in Table~\ref{table_energy}. As shown in Table~\ref{table_energy}, the advantages of M3D 2T-GC memory over conventional 2D SRAM go beyond the footprint reduction afforded by 3D stacking. It also yields a substantial reduction in access energy at identical capacities. This positions M3D 2T-GC as a compelling candidate for future scaled-up, ultra-large on-chip memory systems.

\subsection{Validation}

\begin{figure}[t]
\centerline{\includegraphics[width=19.0pc]{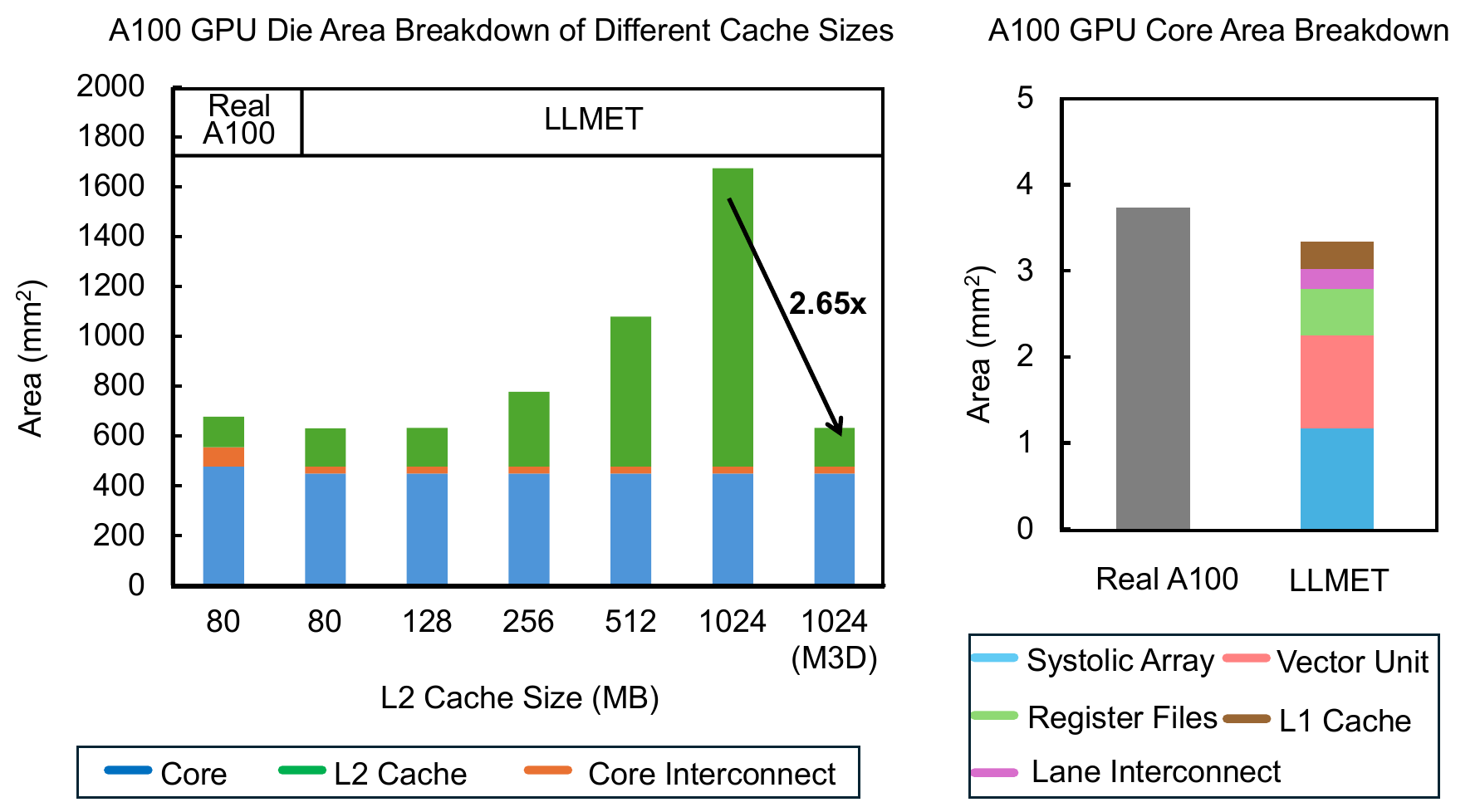}}
\captionsetup{skip=5pt}
\caption{A100 GPU die area breakdown of different cache sizes and corresponding core area breakdown.}\vspace*{-10pt}
\label{eval4}
\end{figure}

The front-end simulator is built upon a widely validated LLM inference framework \cite{llmcompass} calibrated to the NVIDIA A100 GPU. For back-end PPA modeling, circuit blocks are derived from ASAP7 RTL synthesis \cite{asap7}, NS-Cache \cite{nscache}, and NeuroSim \cite{neurosim}. Figure \ref{eval4} compares LLMET area estimates with reported A100 data from \cite{llmcompass,a100datasheet}. The error for both total die area and core area is within 7\%.
Here, all cache capacities are implemented in 2D except the last column. Scaling 2D cache from 40MB to 1GB increases total chip area dramatically. In contrast, M3D integration keeps the area overhead of a 1GB cache within 23\% relative to the 40MB baseline.

\subsection{Server inference (A100, Llama-70B)}

\begin{figure}[t]
\centerline{\includegraphics[width=19.5pc]{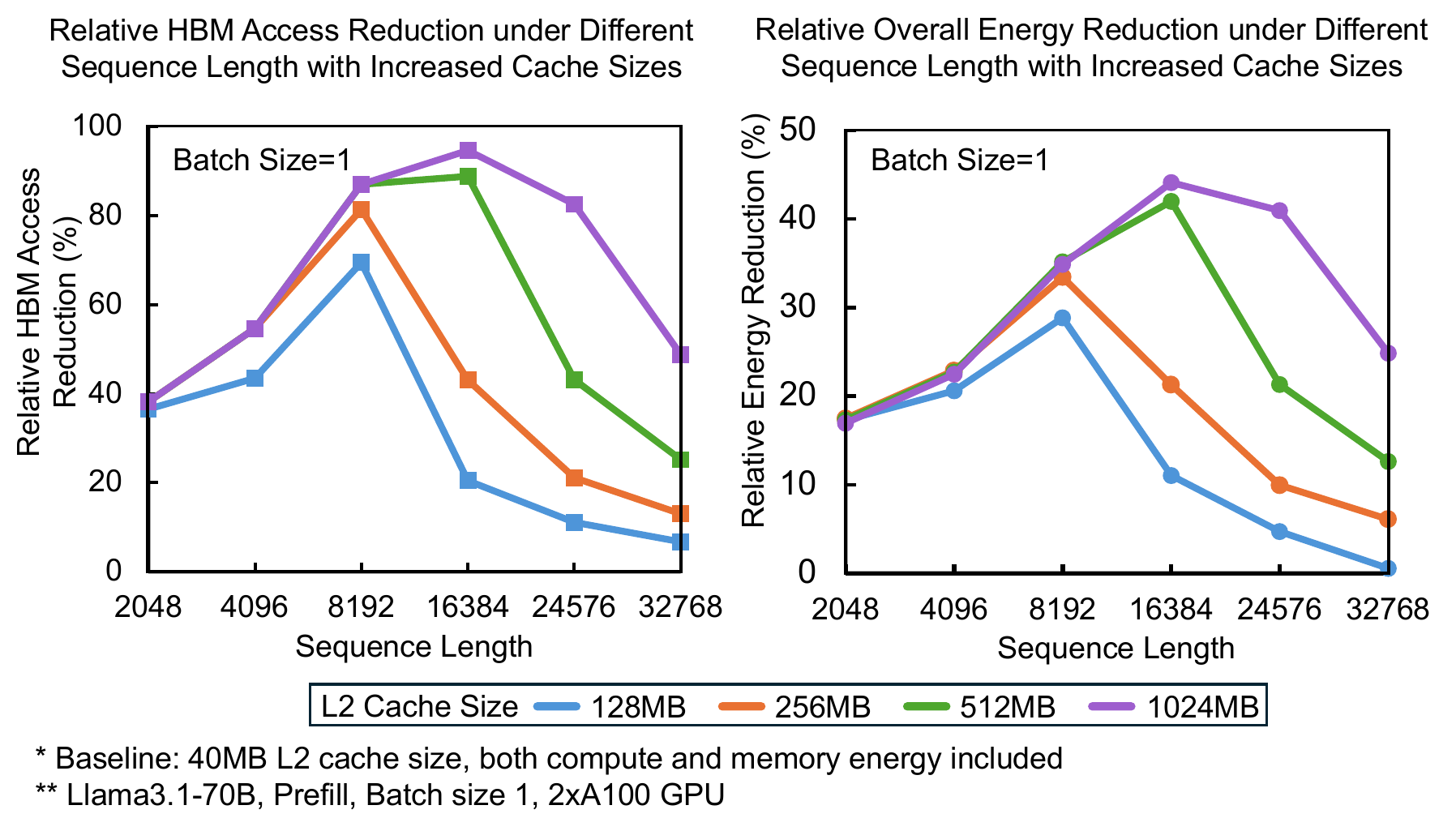}}
\captionsetup{skip=5pt}
\caption{A100: Relative HBM access and overall energy reduction under different sequence length and cache sizes.}\vspace*{-10pt}
\label{eval1}
\end{figure}

We first examine the impact of cache capacity under different sequence lengths. As shown in Figure \ref{eval1}, three observations emerge for the sequence length ranging from 2k to 32k. First, increasing on-chip cache capacity reduces HBM accesses by up to 95\% and total energy by 44\%, enabled by higher data reuse through cache-aware mapping and operator fusion. Second, these benefits gradually saturate as sequence length increases due to less reuse opportunities for longer contexts. The saturation point shifts to larger sequence lengths as cache capacity grows. Third, beyond optimal capacity, further cache expansion diminishes energy savings when reuse gains and HBM access reductions are limited.  

\begin{figure}[t]
\centerline{\includegraphics[width=19.0pc]{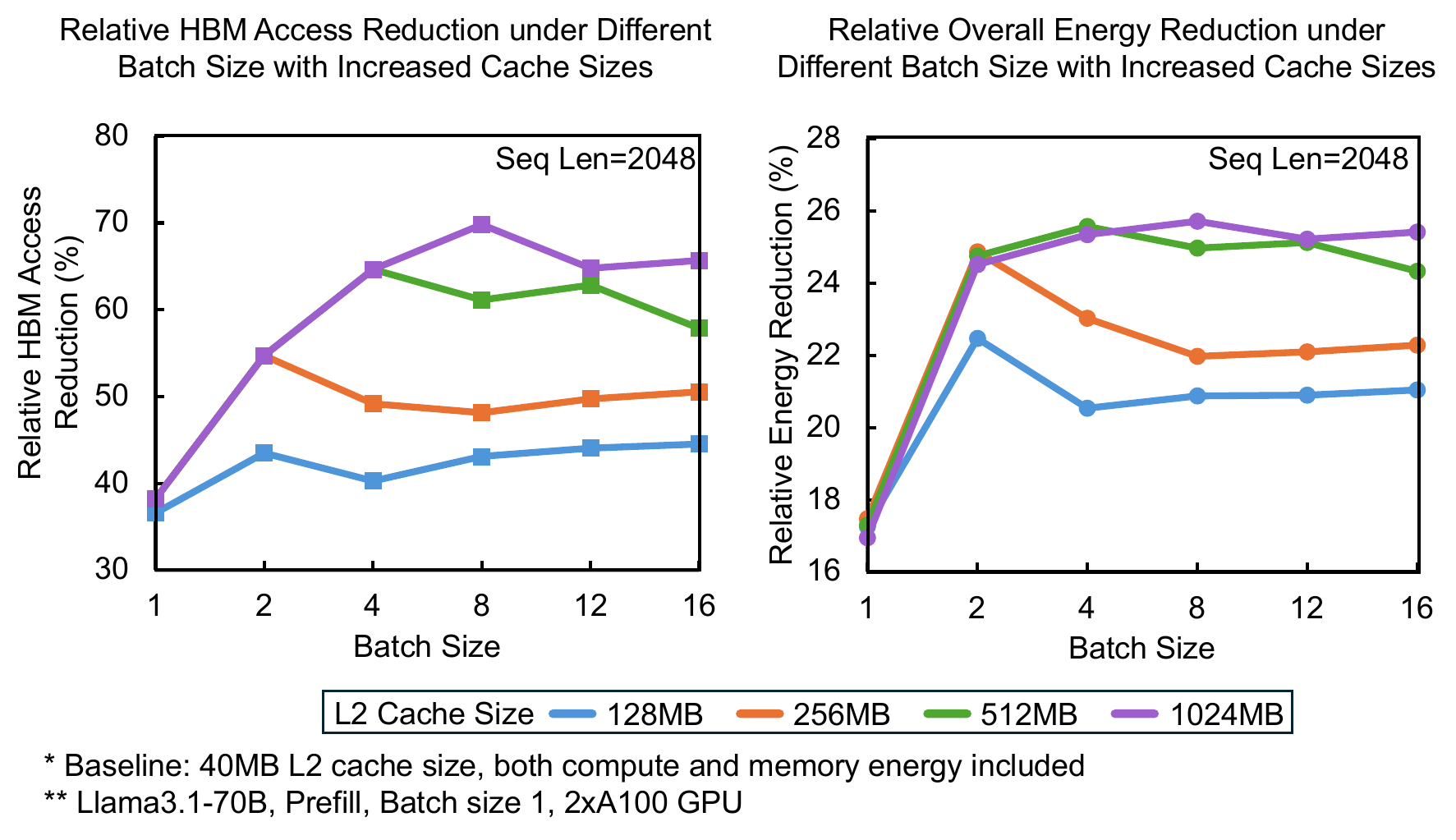}}
\captionsetup{skip=5pt}
\caption{A100: Relative HBM access and overall energy reduction under different batch sizes and cache sizes.}\vspace*{-15pt}
\label{eval2}
\end{figure}

\begin{figure}[t]
\centerline{\includegraphics[width=19.5pc]{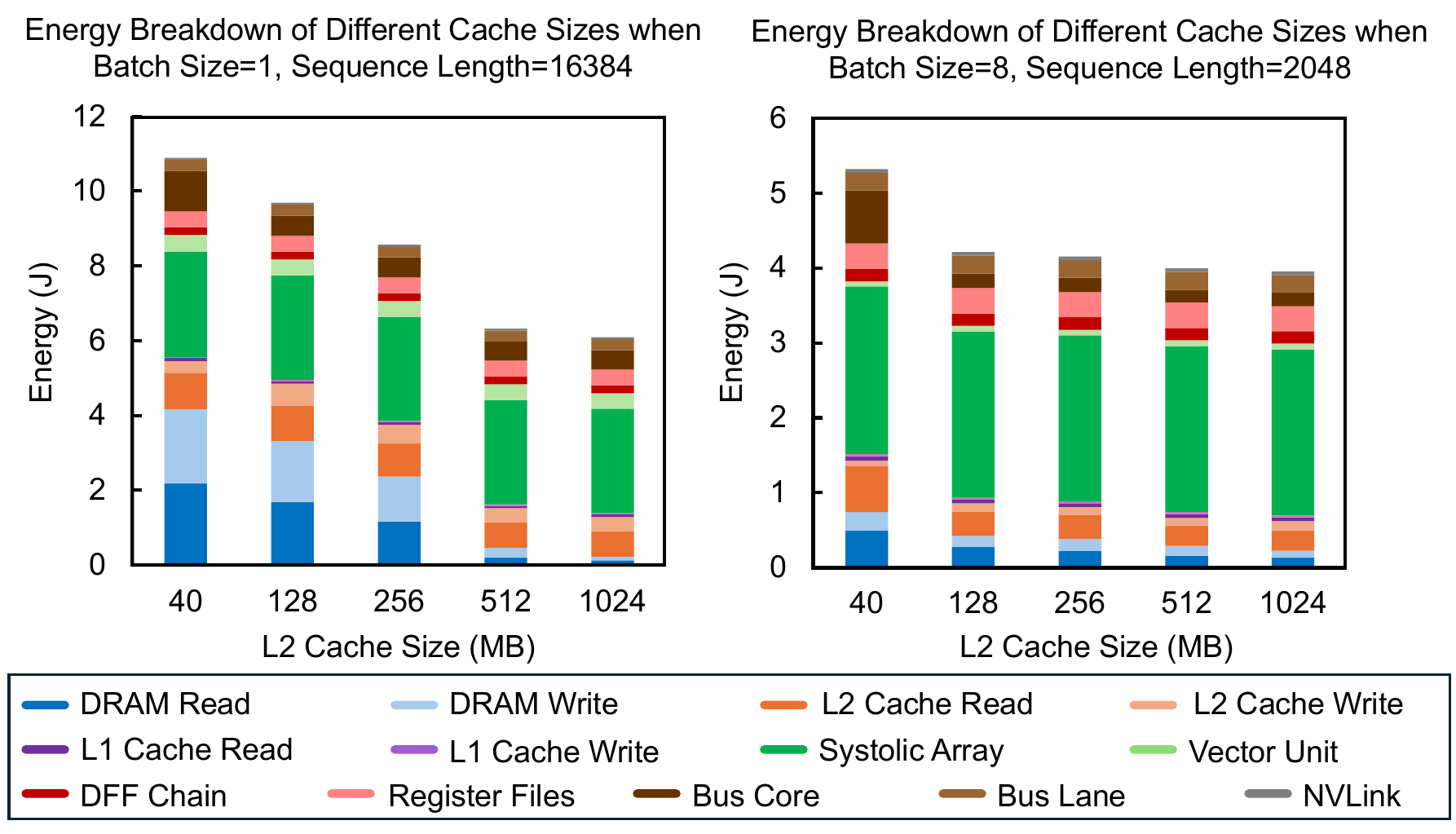}}
\captionsetup{skip=5pt}
\caption{Llama3.1 70B prefill energy breakdown (one layer) of different cache sizes under two batch size and sequence length pairs on two A100 GPUs.}\vspace*{-15pt}
\label{eval3}
\end{figure}

Next, we evaluate different batch sizes at a fixed 2K sequence length. Figure \ref{eval2} shows that HBM traffic reduction and energy benefits saturate faster with batch size than with sequence length. One reason is that attention at 2K already achieves high reuse with modest cache sizes (e.g., 40MB–128MB), so larger batches provide limited additional memory benefit because attention is processed batch-by-batch. In addition, larger batch sizes increase compute energy, reducing overall savings from lower memory traffic. 
We additionally observe that HBM traffic fluctuates for batch sizes 2–8. This occurs because some FFN layers under 128MB and 256MB caches still rely on heuristic mapping, so the selected tiling may not always minimize traffic under limited search space.

To better understand this, Figure \ref{eval3} presents energy breakdowns for two representative workloads: batch-1 with 16K input and batch-8 with 2K input. First, larger caches reduce HBM access energy much more effectively for the long-context batch-1 case. Second, compute energy dominates total energy for batch-8. Together, these effects explain the smaller gains at high batch size.

\begin{figure}[t]
\centerline{\includegraphics[width=19.0pc]{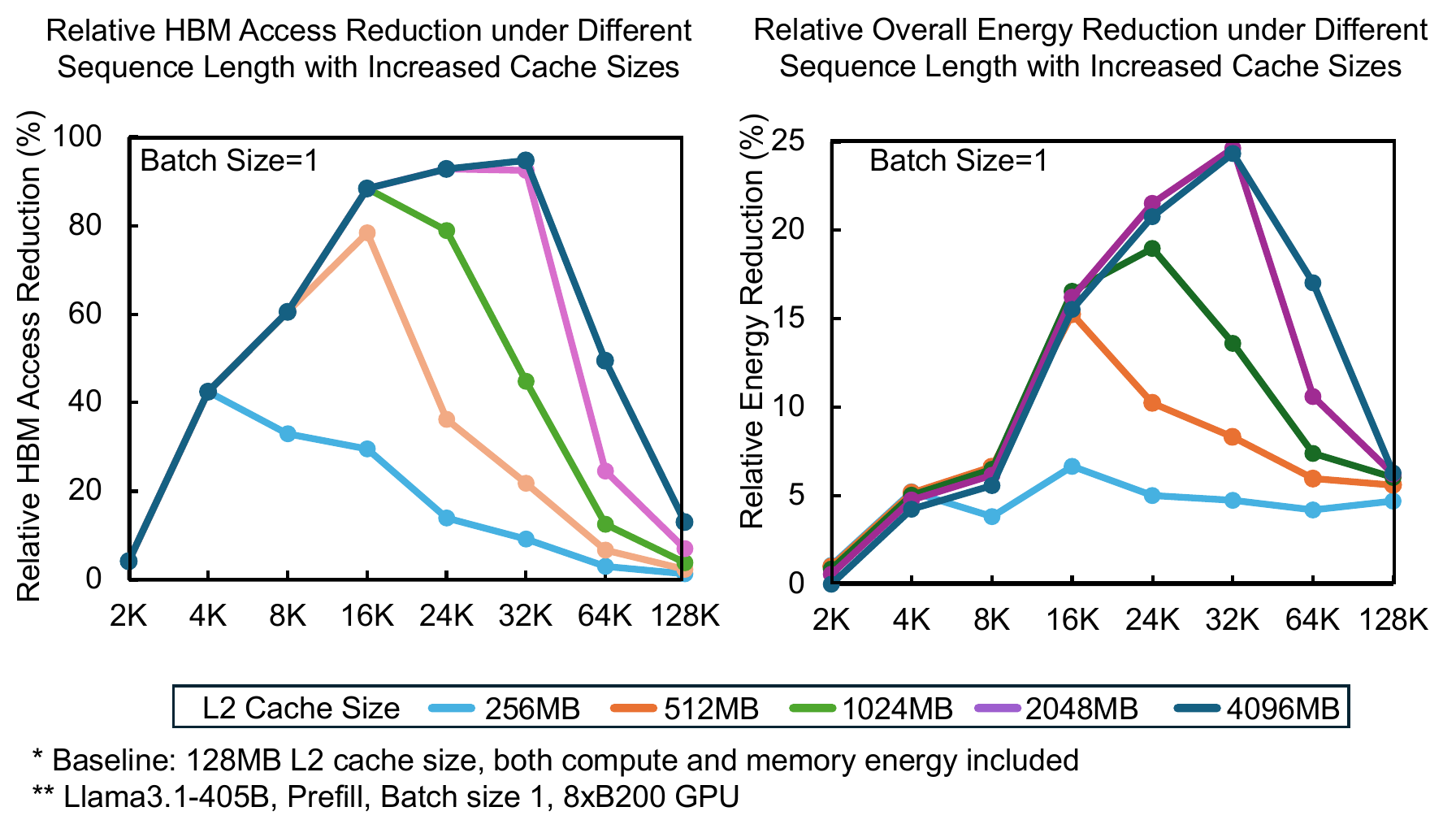}}
\captionsetup{skip=5pt}
\caption{B200-like: relative HBM access and overall energy reduction under different sequence length and cache sizes.}\vspace*{-15pt}
\label{eval5}
\end{figure}

\subsection{Technology Scaling (B200-like, Llama-405B)}
We extend our evaluation from A100 GPUs to a more advanced B200-like platform to assess the impact of large on-chip caches on HBM access reduction and energy efficiency for Llama 3.1 405B. As illustrated in Figure \ref{eval5}, we observe a maximum HBM access reduction of 95\% and peak energy savings of 24\% at a 32K input length. While the scaling trends mirror those of the A100, the peak benefit shifts toward longer sequences as the model and hardware scales up. Notably, excessively large caches introduce access energy overhead, which can saturate net energy gains for long sequences, large batch sizes, or large models. Our results suggest that for sequence lengths between 16K and 64K, a 2GB to 4GB cache is optimal for minimizing prefill energy. Conversely, for sequences shorter than 16K, a 256MB to 512MB cache provides sufficient savings, with further scaling yielding diminishing returns.

The observations in the above two subsections indicate that \textbf{(1) for server inference, prefill phase benefits significantly from larger on-chip cache}; \textbf{(2) optimal cache capacity for maximum energy efficiency shifts across different workloads. } 

\subsection{Edge inference (Jetson-like, Llama-3.2 1B)}

We evaluate edge inference using Llama 3.2 1B quantized to INT4~\cite{gptq, awq} ($\approx$486~MB of weights) on Jetson Orin NX-class accelerators. Four workloads with input/output lengths from 256/128 to 4096/512 tokens are evaluated across L2 capacities from 8~MB to 1~GB. Because edge accelerators are strongly area-constrained, we treat 256~MB as the practical upper range for near-term edge designs, while the 512~MB--1~GB points are included as an upper-bound sensitivity study to quantify the benefit of full on-chip model residency.

\begin{figure}[t]
\centerline{\includegraphics[width=19.5pc]{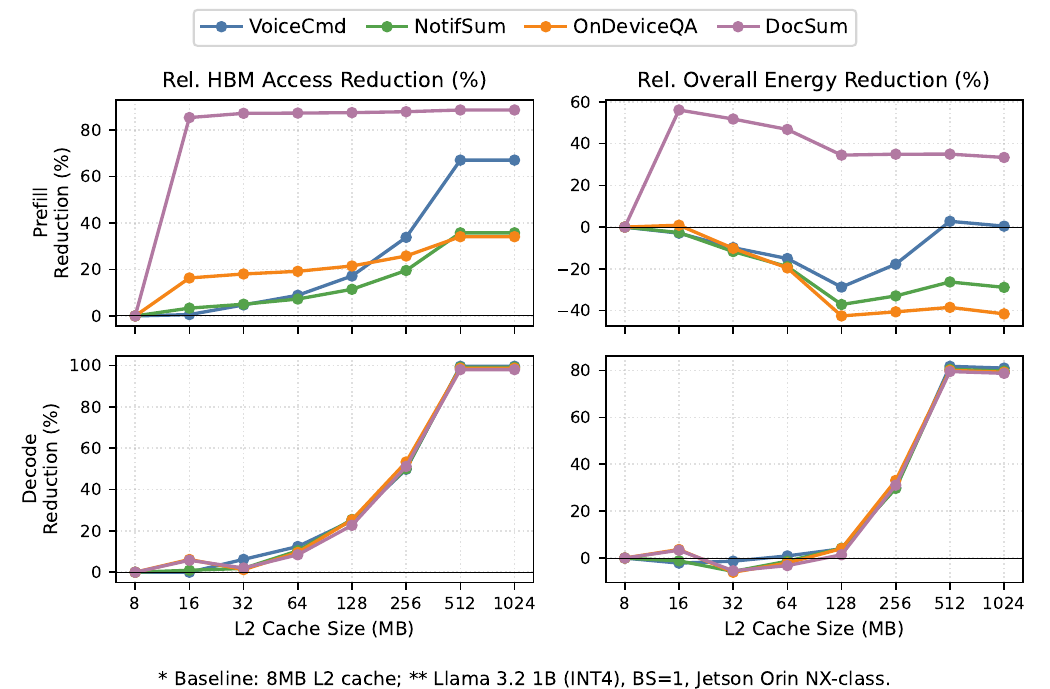}}
\captionsetup{skip=5pt}
\caption{Edge accelerator: relative DRAM access and overall energy reduction across L2 cache sizes for four Llama 3.2 1B INT4 workloads. Top row: prefill; bottom row: decode.}\vspace*{-15pt}
\label{edge_eval1}
\end{figure}

Figure~\ref{edge_eval1} reports off-chip DRAM access and total-energy reductions relative to the 8~MB baseline. During prefill, increasing L2 capacity can reduce DRAM accesses, especially for the longest-context workload (Doc Summary, 4K tokens). However, this reduction does not consistently translate into total-energy savings. For short-prompt edge workloads, the larger on-chip memory introduces higher L2 read/write energy while exposing limited reuse, causing the total prefill energy to remain flat or even increase. As a result, prefill benefits are workload-dependent and should not be viewed as the primary motivation for large edge-side caches; only the long-context Doc Summary workload achieves substantial positive prefill savings.

Decode shows a clearer but more capacity-sensitive trend. Within the practical 256~MB range, larger L2 caches reduce off-chip weight traffic and provide moderate decode-energy savings, but the model still cannot be fully resident on chip. Once the L2 capacity exceeds the quantized model footprint ($\approx$512~MB), the weights become fully resident, DRAM accesses drop by more than 90\%, and decode energy is reduced by 75--80\% at 1~GB across all workloads. These 512~MB--1~GB results therefore represent an aspirational upper bound rather than a near-term edge-cache target.

\begin{figure}[t]
\centerline{\includegraphics[width=19.0pc]{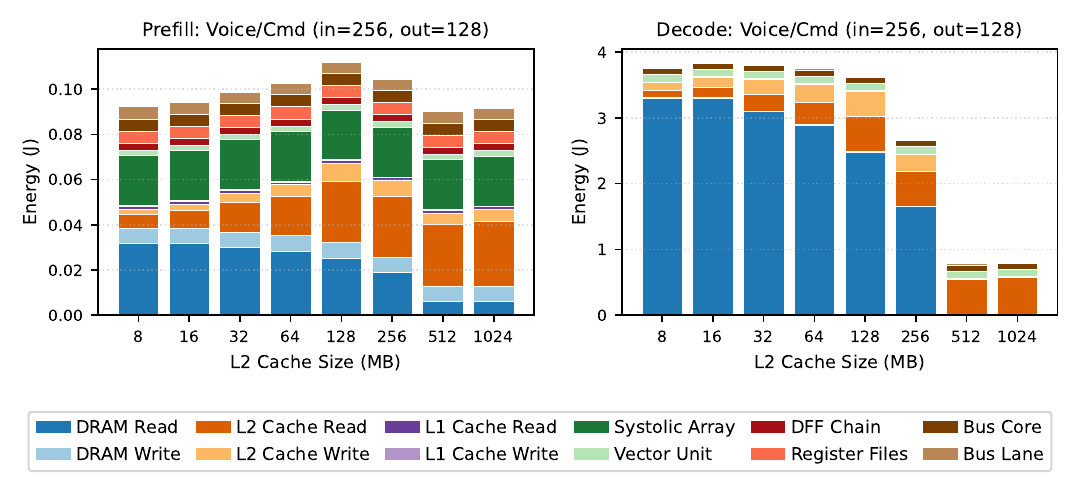}}
\captionsetup{skip=5pt}
\caption{Edge accelerator: per-component energy breakdown (one layer) of the Voice/Cmd Llama 3.2 1B INT4 workload across L2 cache sizes. Left: prefill; right: decode.}\vspace*{-15pt}
\label{edge_eval2}
\end{figure}

Figure~\ref{edge_eval2} further decomposes the energy of the representative Voice/Cmd workload. In decode, DRAM-read energy dominates at small cache capacities and decreases as more model weights are retained on chip, becoming nearly eliminated only when the cache reaches the 512~MB--1~GB residency regime. In contrast, during prefill, the increase in L2 access energy can offset or exceed the reduction in DRAM access energy, explaining why larger caches do not always improve total prefill energy for short edge workloads.

In summary, edge inference exposes a different design tradeoff from server-side prefill. Under realistic edge area constraints, moderately larger caches up to 256~MB can reduce off-chip traffic and improve decode efficiency, but full model residency requires substantially larger capacity and should be interpreted as an upper-bound design point. Therefore, emerging large on-chip memories are most promising for edge decode, while their prefill benefit is limited to long-context workloads with sufficient reuse.

\section{Design Insights and Directions}
\label{sec:insights}

Our cross-layer analysis yields concrete guidelines for the memory systems of future LLM and AI accelerators. Returning to the motivating question of this work---whether emerging M3D on-chip memories can enable energy-efficient LLM serving---our answer is a \emph{qualified yes}: they can, but only within a workload- and technology-dependent regime that the following insights make explicit.

\noindent\textbf{Insight 1: Leverage larger cache for energy savings, up to the phase-specific working-set knee.} Across the server (Figs.~\ref{eval1},~\ref{eval5}) platform, scaling up the on-chip cache capacity yields significant energy benefits for the prefill. However, prefill energy reduction degrades or saturates once cache capacity exceeds a workload's reusable working set. Each workload therefore has an optimal capacity: 256MB--1GB covers 16K--64K contexts for Llama 3.1 70B on A100, while 2--4GB is optimal for Llama 3.1 405B on a B200-like platform and 256--512MB suffices below 16K. Importantly, a larger cache not only drives energy savings in phase-aggregated systems, but also delivers a more profound energy advantage to today's phase-disaggregated architectures. To maximize prefill energy efficiency across different token lengths, platforms should be designed with application- and workload-specific configurability.

\noindent\textbf{Insight 2: The memory technology's access energy sets the break-even point.} Large on-chip memories are useful only when their access energy remains far below HBM energy ($\approx$5.7--6.6~pJ/bit). M3D 2T-GC ($\approx$0.8--1.1~pJ/bit) preserves this margin at GB scale, whereas technologies approaching HBM access energy lose the benefit. Area is equally critical: compared to planar SRAM, M3D memory offers a highly compact footprint with much smaller routing access energy penalties when scaled from MB to GB capacity. Thus, M3D's key value is making the energy-optimal capacity physically affordable on chip.

\noindent\textbf{Insight 3: For edge, full model residency is beneficial but aggressive, and on-chip cache sizing should optimize end-to-end phase composition.}
Edge serving exposes a cross-phase tension absent in server prefill: as on-chip capacity grows (Figs.~\ref{edge_eval1},~\ref{edge_eval2}), prefill energy stays flat or rises slightly for short prompts---the added L2 access energy is not amortized by enough reuse---while decode energy falls steadily as more weights become resident on chip. Fitting the full quantized model on chip (e.g., $\approx$486MB for Llama 3.2 1B INT4) eliminates most decode weight reloads, cutting DRAM traffic by over 90\% and decode energy by 75--80\%, but this capacity is aggressive for area-constrained edge silicon and should be viewed as an upper-bound target; practical capacities up to 256MB still help by retaining part of the model on chip. Because the two phases respond oppositely, the net benefit hinges on a workload's output-to-input ratio, so edge cache sizing should be driven by the deployment's input/output length distribution rather than by optimizing either phase in isolation.

\vspace{-5pt}
\section{Conclusion and Future Work}

This work first presents LLMET, a validated cross-layer framework for co-designing LLM serving system and memory technologies. We then evaluate such technologies' impact on LLM inference across platforms and provide concrete system design insights. Our findings position emerging memory as a cross-layer design knob for future energy-efficient LLM accelerators and highlight directions for embedded memory research in emerging AI workloads.



\end{document}